%% file: FASGEP_2016.tex
\documentclass[11pt,a4paper]{article}

\input{packages}

\input{macro}

\usepackage{authblk}

\title{Counting Constraints in Flat Array Fragments}
\author[1]{Francesco Alberti}
\author[2]{Silvio Ghilardi}
\author[2]{Elena Pagani}

\affil[1]{ Fondazione S. Raffaele, Milano, Italy}
\affil[2]{Universit\`a degli Studi di Milano, Milano, Italy}

\newtheorem{theorem}{Theorem}
\newtheorem{definition}{Definition}
\newtheorem{lemma}{Lemma}

\begin{document}
%
%\frontmatter          % for the preliminaries
%
%\pagestyle{headings}  % switches on printing of running heads
%JhalaM07
%\mainmatter              % start of the contributions
%

%
%
\maketitle              % typeset the title of the contribution
\setcounter{footnote}{0}

\begin{abstract}
We identify a fragment of Presburger arithmetic enriched with free function symbols and cardinality constraints for interpreted sets, which is amenable to automated analysis. We establish decidability and complexity results for such a fragment and
we implement our algorithms. The experiments run in discharging proof obligations
coming from invariant checking and bounded model-checking benchmarks
show the practical feasibility of our decision procedure.
%
% We identify a fragment of Presburger arithmetic enriched with free function symbols and cardinality constraints for interpreted sets, which is amenable to automated analysis. 
% We establish decidability and complexity results, and we experiment practical feasibility of our algorithms with a prototype which is shown to be
% able to discharge proof obligations coming from invariant checking and bounded model-checking benchmarks.
\end{abstract}

\section{Introduction} 

Enriching logic formalisms with counting capabilities is an important task in view of the needs of many application areas, ranging from database theory to formal verification. 
Such enrichments have been designed both in the description logics area and in the area of Satisfiability Modulo Theories (SMT), where some of the most important recent achievements 
were decidability
and  complexity bounds for BAPA~\cite{KunkakCADE20} - the enrichment of Presburger arithmetic with the ability of talking about finite sets and their cardinalities. 
As pointed out in~\cite{KunkakJAR}, BAPA constraints can be used for program analysis and verification by expressing data structure invariants, simulations between program fragments or termination conditions.
The analysis of BAPA constraints was successfully extended also to formalisms encompassing multisets~\cite{KunkakVMCAI08} as well as direct/inverse images 
%silvio lascio 'along' (='lungo') è standard
along relations and functions~\cite{KunkakVMCAI10}.

A limitation of BAPA and its extensions lies in the fact that only uninterpreted symbols (for sets, relations, functions, etc.) are allowed. On the other hand, 
it is well-known that a different logical formalism, namely
unary counting quantifiers, can be used  in order to reason about the cardinality of definable (i.e. of interpreted) sets.  
Unary counting quantifiers can be added to Presburger arithmetic without compromising  decidability, see~\cite{schweikhart}, 
however they might be quite problematic if combined in an unlimited way with free function symbols.
In this paper, we investigate the extension of Presburger arithmetic including both counting quantifiers and uninterpreted function symbols, and we isolate fragments where we can achieve decidability and in some cases also relatively good complexity bounds. The key ingredient to isolate such fragments is the notion of flatness: roughly, in a flat formula, subterms of the kind $a(t)$ (where $a$ is a free function symbol) can occur only if $t$ is a variable. By itself, this naif flatness requirement is useless (any formula can match it to the price of introducing extra quantified variables); 
in order to make it effective, further 
 syntactic restrictions should be incorporated in it, as witnessed in~\cite{AlbertiGS14}. This is what we are going to do in  this paper, where suitable notions of `flat' and `simple flat' \formulae are introduced
 in the rich context of Presburger arithmetic enriched with free function symbols  and with unary counting quantifiers (we use free function symbols to model arrays, see below).

The fragments we design are all obviously more expressive than BAPA, but they do not come from pure logic motivations, on the contrary they are suggested by an emerging  application area, 
namely the area of verification of fault-tolerant distributed systems.
Such systems (see~\cite{zufferey} for a good account)  are modeled as partially synchronous systems, where a finite number of identical
 processes operate  in lock-step (in each \textit{round} they send messages,
receive messages, and update their local state depending on the local state at the beginning of
the round and the received messages). Messages can be lost, processes may omit to perform some tasks or also behave in a malicious way; for these reasons, 
the fact that some actions  are enabled or not, and the correctness of the algorithms themselves,  are subject to threshold conditions saying for instance  that
some qualified majority of processes are in a certain status or behave in a non-faulty way. Verifications tasks thus have to handle  cardinality constraints
of the kind studied in this paper (the reader interested in full formalization examples can directly go to Section~\ref{sec:experiments}).

The paper is organized as follows: we first present basic syntax (Section~\ref{sec:preliminaries}), then decidability (Section~\ref{sec:sat})  and complexity (Section~\ref{sec:tractable}) results; 
experiments with our prototypical implementation are supplied in Section~\ref{sec:experiments}, and  Section~\ref{sec:conclusions} concludes the work.

\section{Preliminaries}\label{sec:preliminaries}

We work within Presburger arithmetic enriched with free function symbols and cardinality constraints. This is a rather expressive logic, whose syntax is summarized in Figure~\ref{fig:syntax}.
Terms and \formulae are interpreted in the natural way over the domain of integers $\mathbb Z$; as a consequence, satisfiability of a formula $\phi$ means that it is 
possible to assign values to parameters, free variables and array-ids so as to make  $\phi$ true in $\mathbb Z$ 
(validity of $\phi$ means that $\neg \phi$ is not satisfiable, equivalence of $\phi$ and $\psi$ means that $\phi\leftrightarrow \psi$ is valid, etc.). We nevertheless implicitly assume few constraints (to be explained below) 
about our intended semantics.

To denote integer numbers, we have (besides variables and numerals) also
parameters: the latter denote unspecified integers. 
%(expressing e.g. in the intended applications the number of  faulty processes); 
Among parameters, we always include a specific parameter (named $N$) 
% avoid repetition denote-denoting
%denoting
identifying
the dimension of the system - 
alias the length of our arrays: in other words, it is assumed that for all array identifiers $a\in Arr$, the value $a(x)$ is  conventional 
(say, zero) outside the interval $[0, N)=\{ n\in \mathbb Z\mid 0\leq n < N\}$.
Although binary free function symbols are quite useful in some applications, in this paper we prefer not
to deal with them.
The operator $\sharp\,\{ x \mid  \phi\}$ \emph{indicates the cardinality of the 
finite set formed by the $x\in [0, N)$ such that $\phi(x)$ holds}.

Notice that the cardinality constraint operator $\sharp\,\{ x \mid -\}$, as 
well the  quantifier $\exists x$, 
binds the variable $x$;   
below, we indicate with $\psi(\ux)$ (resp. $t(\ux)$) the fact that the formula $\psi$ (the term $t$) has free individual variables included in the list $\ux$. 
%Substitution of the free occurrences of a variable $x$ by a term $t$ 
%in a formula $\psi$ is indicated with
%$\psi(t/x)$; 
When we speak of a substitution, we always mean `substitution without capture', meaning that,
when we replace the free occurrences of a variable $x$ with a term $u$ in a formula
$\phi$ or in a term  $t$, 
the term $u$ should not contain free variables that might be located inside  the scope of a 
binder for them once the substitution is performed; the result of the substitution is denoted with $\phi(u/x)$ and $t(u/x)$.

The logic of Figure~\ref{fig:syntax} is far from being tractable, because even
the combination of  free function symbols and Presburger arithmetic lands in a highly undecidable class~\cite{halpern}. We are looking for a mild fragment, nevertheless
sufficiently expressive  for our intended applications. These applications mostly come from verification tasks, like bounded model checking or invariant checking. 
Our aim is to design a decidable fragment (so as to be able not only to produce certifications, but also to find bugs) with some minimal closure properties; from this point of view, 
notice that for bounded model checking closure under 
conjunctions is sufficient, but for invariant checking we need also closure under negations in order to discharge entailments.

%Before formally introducing it, let us point out the model checking tasks we should be able to perform.

\begin{figure}[t]
\centering
{\scriptsize
\begin{tabular}{lll}
$0, 1, \dots $ 
  &$\in \mathbb Z$& numerals (numeric constants) \\ 
$x, y, z, \dots $ 
  &$\in Var$& individual variables \\ 
   $ M, N, \dots $ 
  &$\in Par$& parameters (free constants)
  \\
   $a,b, \dots $ 
   &$\in Arr$& array ids
   (free unary 
   \\&& function symbols)
  \\
  $t, u, \dots~~~::=~~$
  &$n \;\vert\; M \;\vert\; x \;\vert\; t+t\; \vert\; -\!t\; \vert\; a(t) \;\vert\; \sharp\{ x\mid \phi\}~$
  & terms
  \\
  $A, B,\dots~::=~$
  & $t<t \;\vert\; t=t \;\vert\; t\equiv_n t~$ & atoms
  \\
  $\phi, \psi, \dots~~::=~~ $
  &$A\; \vert \; \phi\wedge \phi\; \vert\; \neg \phi\; \vert \; \exists x\, \phi~$ &formulae
\end{tabular}
}
\caption{Syntax\label{fig:syntax}}
\end{figure}

\subsection{Flat  \formulae}
We now introduce some useful subclasses
of the \formulae built up according to the grammar of Figure~\ref{fig:syntax}.
 %Let us call
 \begin{description}
  \item[-] \emph{Arithmetic \formulae}: these are built up from the  grammar of Figure~\ref{fig:syntax} without using
 neither array-ids nor cardinality constraint operators; we use letter $\alpha, \beta, \dots$ for
 arithmetic \formulae. Recall that, according to the well-known quantifier elimination result, arithmetic \formulae are equivalent 
 to \emph{quantifier-free} arithmetic \formulae. 
  \item[-] \emph{Constraint \formulae}: these are built up from the  grammar of Figure~\ref{fig:syntax} without using
 array-ids.
  \item[-] \emph{Basic \formulae}: these are obtained from an arithmetic formula by simultaneously replacing some free variables by terms of the kind $a(y)$, \emph{where $y$ is a variable
  and $a$ an array-id}. When we need to display full information, we may use the notation $\alpha(\uy, \ta(\uy))$  to indicate basic \formulae. By this notation, we mean that $\uy=y_1, \dots, y_n$ are variables,
$\ta=a_1, \dots, a_s$ are array-ids and that $\alpha(\uy, \ta(\uy))$ is obtained from an arithmetic formula $\alpha(\ux, \uz)$ (where $\uz=z_{11}, \dots, z_{sn}$)
by replacing $z_{ij}$ with $a_i(y_j)$ ($i=1, \dots, s$ and $j=1,\dots, n$). 
%When we write $\alpha(\ux, \ta(\ux))$, we make the conventions $\alpha(\ux, \ta(\ux))$ that the free individual variables occurring in 
%$\alpha(\ux, \ta(\ux))$ are included in the list $\ux$ and the array-ids occurring in $\alpha(\ux, \ta(\ux))$ are included in the list $\ta$. 
  \item[-] \emph{Flat \formulae}: these are recursively defined as follows (i) basic \formulae are %acyclic
  flat \formulae; (ii) if $\phi$ is 
  %an acyclic 
  a flat  formula, $\beta$ is a basic formula, $z$ and $x$ are variables, 
  then $\phi(\sharp\,\{x\!\mid\! \beta\}\,/\,z)$ is 
  %an acyclic 
  a flat formula.\footnote{ If we want to emphasize the way 
  %silvio little rewording
  the basic formula
  $\beta$ is built up, 
  following the above conventions,
  we may write it as 
  %above as 
  $\beta(x, \uy,\ta(x), \ta(\uy))$; here, supposing that  
  $\ta$ is $a_1, \dots, a_s$, since $x$ is a singleton, the tuple
  $\ta(x)$ is $a_1(x), \dots, a_s(x)$.
  }
 \end{description}

Notice that all the above classes are closed under Boolean operations (in particular, under negations). 
The following result is proved in~\cite{schweikhart} (see also Appendix~\ref{app:counting}):

\begin{theorem}\label{thm:NS}
 For every constraint formula one can compute an arithmetic formula equivalent to it.
\end{theorem}

%[Examples needed]

\section{Satisfiability for flat \formulae}\label{sec:sat}

We shall show that flat \formulae are decidable for satisfiability. In fact, we shall show decidability of the slightly
larger class covered by the following

\begin{definition}
Extended flat \formulae (briefly, E-flat \formulae) are \formulae of the kind
   \begin{equation}\label{eq:flatdef}
\exists \uz.~\alpha~ \wedge~ \sharp\{x\mid \beta_1\} =z_1~\wedge \cdots \wedge~ \sharp\{x\mid \beta_K\} =z_K
\end{equation}
%silvio  ho messo qualchje precisazione in più
  where $\uz=z_1, \dots, z_K$ and $\alpha,\beta_1, \dots, \beta_K$ are basic \formulae and $x$ does not occur in $\alpha$.
\end{definition}

Notice that 
%nothing prevents 
$\alpha$ and the $\beta_j$ in~\eqref{eq:flatdef} above 
%from   containing 
may contain
further free variables $\uy$ (besides $\uz$) as well as the terms  $\ta(\uy)$ and $\ta(\uz)$; the $\beta_j$ may contain occurrences of $x$ and of $\ta(x)$.

 That 
 %acyclic 
 flat \formulae are also E-flat  can be seen 
as follows: due to the fact that our substitutions avoid captures, we can use equivalences like 
 $\phi(t/z)\leftrightarrow \exists z\,(t=z \wedge \phi)$  
 in order to abstract out the terms $t:=\sharp\,\{x\!\mid\! \alpha\}$ occurring in the recursive construction of a %
 %acyclic 
 flat  formula
 $\phi$. By repeating this  
linear time transformation, we end up in a formula of the kind~\eqref{eq:flatdef}. However, not all E-flat \formulae are %acyclic 
flat because
the dependency graph associated to~\eqref{eq:flatdef}  might not be acyclic (the graph we are talking about has the $z_j$ as nodes and has
an arc $z_j\to z_i$ when $z_i$ occurs in $\beta_j$). The above conversion of a flat formula into a formula of the form~\eqref{eq:flatdef} on the other hand produces an E-flat formula whose associated graph is acyclic.

We  prove a technical lemma showing how we can manipulate
E-flat \formulae without loss of generality.
Formulae 
$\varphi_1, \dots, \varphi_K$ are said to be a partition iff the \formulae $\bigvee_{l=1}^K \varphi_l$ and $\neg(\varphi_l\wedge \varphi_h)$ (for $h\neq l$)
are valid. %(i.e. true in our intended structures for every valuation of the free variables). 
Recall that the existential closure of a formula is the sentence obtained by prefixing it with a string of existential quantifiers binding all variables having a free occurrence in it.

\begin{lemma}\label{lemma:nf}
 The existential closure of an E-flat  formula  is equivalent to a sentence of the kind 
 \begin{equation}\label{eq:card}
\exists \uz\;\exists \uy.~~\alpha(\uy, \uz)~ \wedge~ \sharp\{x\mid \beta_1(x, \ta(x), \uy, \uz)\} =z_1~\wedge \cdots \wedge~ \sharp\{x\mid \beta_K(x, \ta(x),\uy, \uz )\} =z_K
\end{equation}
 where $\uy$ and $\uz:=z_1, \dots, z_K$ are  variables, $\alpha$ is arithmetical, and the \formulae $\beta_1, \dots, \beta_K$  are basic and form a partition.%\footnote{Notice that
 %$\alpha$ is an arithmetic formula and the $\beta_l$ are basic \formulae.}
\end{lemma}

\noindent
\textit{Proof.} The differences between (the matrices of)~\eqref{eq:card} and~\eqref{eq:flatdef} are twofold: first in~\eqref{eq:card}, the $\beta_l$ form a partition and, second, in~\eqref{eq:flatdef} the terms $a_s(y_i)$ and $a_s(z_h)$ 
(for $a_s\in \ta$ and $y_i\in \uy$, $z_h\in \uz$) may occur in $\alpha$ 
%(which is assumed to be only basic) 
and in the $\beta_l$.

We may disregard the $a_s(z_h)$ without loss of generality, because we can include them in the $a_s(y_i)$: to this aim, it is sufficient to take a fresh $y$, to add
the conjunct $y=z_h$ to $\alpha$ and to replace everywhere $a_s(z_h)$ by $a_s(y)$.
  In order 
%to get~\eqref{eq:card}, we need  
to  eliminate also a term like $a_s(y_i)$, we make a guess and distinguish 
the case where $y_i\geq N$ and the case where $y_i < N$ (formally, `making a guess' means to replace~\eqref{eq:flatdef} with a disjunction - the two disjuncts being obtained by adding to $\alpha$ the case description). According to the semantics conventions we made in Section~\ref{sec:preliminaries}, the first case is trivial
because we can just replace $a_s(y_i)$ by 0. In the other case, we
first take  a fresh variable $u$ and apply  
 the equivalence
$\gamma(\dots a_s(y_j)\dots)\leftrightarrow \exists u\, (a_s(y_j)=u \,\wedge\, \gamma(\dots u\dots))$
(here $\gamma$ is the whole~\eqref{eq:flatdef});  then we replace $a_s(y_j)=u$ by the equivalent formula 
 $\sharp\{x \mid x=y_j \wedge a_s[x]=u\}=1$
and 
finally the latter by $\exists u'\, (u'=1 \wedge\, \sharp\,\{x \mid x=y_j \wedge a_s[x]=u\}=u')$ 
 (the result has the desired shape once we move the new existential  quantifiers in front).

% After these transformations, we get (in linear time) a formula having the shape~\eqref{eq:card}; 
After this, we still need to modify the $\beta_l$ so that they form a
 partition (this further step produces an exponential blow-up). Let  $\psi(\uy)$ be the matrix of a formula of the kind~\eqref{eq:card}, where the $\beta_l$ are 
 not a partition. Let us put $\underline K:=\{1, \dots, K\}$ and let us consider further variables $\uu=\langle u_{\sigma}\rangle_{\sigma}$, for $\sigma\in 2^{\underline K} $.
 Then it is clear that the existential closure of $\psi$ is equivalent to the formula obtained by prefixing the existential quantifiers $\exists \uu \,\exists \uz$ to the  formula
 \begin{equation}\label{eq:card0}
\left(\alpha~ \wedge~ \bigwedge_{l=1}^K  z_l = \sum_{\sigma\in 2^{\underline{K}},\;\sigma(l)=1} u_{\sigma}\right)
\wedge \bigwedge_{\sigma\in 2^{\underline{K}}}\sharp\{x\mid \beta_{\sigma}\} =u_{\sigma}  
\end{equation}
where $\beta_{\sigma};= \bigwedge_{l=1}^K \epsilon_{\sigma(l)} \beta_l$ (here $\epsilon_{\sigma(l)}$ is `$\neg$' if $\sigma(l)=0$, it is 
%silvio
%`$~$'  
a blank space otherwise).
$\hfill\dashv$

\begin{theorem}\label{thm:main}
 Satisfiability of E-flat  \formulae is decidable.
\end{theorem}

\noindent
\textit{Proof.} 
 We reduce satisfiability of~\eqref{eq:card} to satisfiability of constraint \formulae
which is decidable by Theorem~\ref{thm:NS}; in detail,
 we show that~\eqref{eq:card} is equisatisfiable with 
the constraint formula below (containing extra free variables $z_{S}, z_{l, S}$):
%, d_{S, \mathcal S}$):
\begin{equation}\label{eq:card1}
  %\begin{aligned}
  %& \gamma \wedge \bigwedge_{\mathcal S\in \wp(\wp(\underline K))} \left( d_{\mathcal S} = \sharp \{ i\mid \bigwedge_{S\in \mathcal S} \exists y \bigwedge_{l\in S} \varphi_l(i,y) \wedge
  % \bigwedge_{S\not\in \mathcal S} \forall y \neg\bigwedge_{l\in S} \varphi_l(i,y)\}\right ) \wedge
  % \\
  % & \wedge \bigwedge_{\mathcal S\in \wp(\wp(\underline K))} \left( d_{\mathcal S} = \sum_{S\in \mathcal S} d_{S, \mathcal S} \right) \wedge \bigwedge_{l=1}^K 
  % \left( d_l = \sum_{\mathcal S\in \wp(\wp(\underline K))} \sum_{S\in \mathcal S, l\in S} d_{S, \mathcal S}\right)
  %\end{aligned}
  \begin{aligned}
  & \alpha \wedge \bigwedge_{S\in \wp(\underline K)} \left( z_{S} = \sharp \{ x\mid \bigwedge_{l\in  S} \;\exists \uu \, \beta_l(x,\uu,\uy,\uz) \wedge
   \bigwedge_{l\not\in S} \forall \uu \neg \beta_l(x,\uu,\uy,\uz)\}\right )  \wedge
   \\
   & \wedge \bigwedge_{ S\in \wp(\underline K)} \left( z_{ S} = \sum_{l\in S} z_{l, S} \right) \wedge \bigwedge_{l=1}^K 
   \left( z_l = \sum_{S\in \wp(\underline K), l\in S} z_{l,S}\right) \wedge  \bigwedge_{ l\in S\in \wp(\underline K)} z_{l,S}\geq 0
  \end{aligned}
\end{equation}
(according to our notations, the basic \formulae
$\beta_l(x, \ta(x), \uy, \uz)$ from~\eqref{eq:card} 
were supposed to be built up from the arithmetic \formulae $\beta_l(x, \uu, \uy, \uz)$ by replacing the variables $\uu= u_1, \dots, u_s$ with 
the terms $\ta(x)= a_1x), \dots, a_s(x)$). 
%the variables $\uu=u_1, \dots u_s$ replace the terms $\ta(x)= a_1(x), \dots, a_s(x)$).

%Proof of equisatisfiability is below.

\emph{Suppose that~\eqref{eq:card1} is satisfiable}. Then there is an assignment $V$ to the  free variables occurring in it so that~\eqref{eq:card1} is true 
in the standard structure of the integers (for simplicity, we use the same name for a free variable and for the integer assigned to it by $V$). If $\ta=a_1, \dots, a_s$,
we need to define $a_s(i)$ for all $s$ and for all $i\in [0, N)$.
For every $l=1, \dots, K$ this must be done in such a way that there are exactly $z_l$ integer numbers  taken from $[0,N)$ satisfying $\beta_l(x, \ta(x),\uy, \uz)$.
%Since the $\varphi_l$ are a partition, 
The interval $[0,N)$ can be partioned by associating with each $i\in [0, N)$ the set  
$i_S\;=\;\{ l\in \underline{K} \mid \exists \uu\;\beta_l(i,\uu, \uy, \uz)$ holds under $V\}$. For every $S\in \wp(\underline{K})$ the number of the $i$ such that $i_S=S$ is $z_S$; for every $l\in S$, pick $z_{l,S}$ among them and, for these selected $i$,  let the $s$-tuple $\ta(i)$ be equal to an $s$-tuple $\uy$ such that $\beta_l(i,\uu, \uy, \uz)$ holds (for this tuple $\uy$, since the $\beta_l$ are a partition, $\beta_h(i,\uu, \uy, \uz)$ does not hold, if $h\neq l$).
Since $z_{ S} = \sum_{l\in S} z_{l, S}$ and since $\sum_S z_S$ is equal to the length of the interval $[0,N)$, the definition of the $\ta$ is
complete. The formula~\eqref{eq:card} is true by construction.

On the other hand \emph{suppose that~\eqref{eq:card} is satisfiable} under an assignment $V$; we need to find $V(z_S)$, $V(z_{l,S})$ 
(we again indicate them simply as $z_S, z_{l,S}$)
so that~\eqref{eq:card1} is true. For $z_S$ there is no choice, since $z_S=\sharp \{ i\mid \bigwedge_{l\in  S} \;\exists \uu \, 
\beta_l(i,\uu,\uy,\uz) \wedge\bigwedge_{l\not\in S} \forall \uu \neg \beta_l(i,\uu,\uy,\uz)\}$ must be true; for $z_{l,S}$, we take it to be the cardinality of the set of the 
$i$ such that $\beta_l(i, \ta(i), \uy, \uz)$ holds under $V$ and $S= \{ h\in \underline{K} \mid \exists \uu\;\beta_h(i,\uu, \uy,\uz)$ holds under $V\}$. In this way, 
for every $S$, the equality $ z_{ S} = \sum_{l\in S} z_{l, S}$ holds and for every $l$, the equality
$z_l = \sum_{S\in \wp(\underline K), l\in S} z_{l,S}$ holds too. Thus the formula~\eqref{eq:card} becomes true under our extended $V$.
 $\hfill\dashv$

\section{A more tractable subcase}\label{sec:tractable}

Thus satisfiability of flat \formulae is decidable; since flat \formulae are closed under Boolean combinations, validity of implications of flat sentences 
is decidable too. This makes
our
%the
result a \emph{complete} algorithm for checking invariants in verification applications.
However, the complexity of the decision procedure is very high:  Lemma~\ref{lemma:nf} introduces an exponential blow-up and other 
exponential blow-ups are introduced by Theorem~\ref{thm:main}
and by the
decision procedure (via quantifier elimination)  from~\cite{schweikhart}. 
Of course, all this might be subject to dramatic optimizations (to be investigated by future reseach); in this paper we show  that there is 
 a much milder 
(and still practically useful) fragment.

\begin{definition}
 Simple flat \formulae are recursively defined as follows: (i) basic \formulae are simple flat \formulae; 
(ii) if $\phi$ is a simple flat formula,
$\beta(\ta(x), \ta(\uy), \uy)$ is a basic formula and $x, z$ are variables, then $\phi(\sharp\{x\!\mid\! \beta\}\,/\,z)$ is a  simple flat formula.
\end{definition}

As an example of a simple flat formula consider the following one 
$$
a'(y)= z \;\wedge\; \sharp\, \{x \mid a'(x)=a(x)\}\geq N\!-\!1\; \wedge\; (\sharp\, \{x \mid a'(x)=a(x)\}<N \to a(y)  \neq z)
$$
expressing that $a'=write(a, y, z)$ (i.e. that the array $a'$ is obtained from $a$ by over-writing $z$ in the entry $y$). 

\begin{definition}
Simple E-flat \formulae are \formulae of the kind 
  \begin{equation}\label{eq:simpleflatdef}
  \begin{aligned}
   \exists \uz.~\alpha(\ta(\uy),\ta(\uz),\uy, \uz)~ \wedge~ \sharp\{x\mid \beta_1(\ta(x),\ta(\uy),\ta(\uz), \uy,\uz)\} =z_1~\wedge \cdots
   \\
   \cdots\wedge~ \sharp\{x\mid \beta_K(\ta(x),\ta(\uy), \ta(\uz), \uy,\uz))\} =z_K~~~~~~
  \end{aligned}
\end{equation}
where $\alpha$ and the $\beta_i$ are basic.
\end{definition}

It is easily seen that (once again) simple flat \formulae are closed under Boolean combinations and that simple flat \formulae are simple E-flat \formulae 
(the converse is not true, for ciclicity of the dependence graph of the $z_i$'s in~\eqref{eq:simpleflatdef}). 

The difference between simple and non simple
flat/E-flat \formulae is that in simple \formulae \emph{the abstraction variable cannot occur outside the read of an array symbol} (in other words,  the $\beta,\beta_i$ from the
above definition are of the kind $\beta_i(\ta(x),\ta(\uy), \ta(\uz), \uy,\uz)$ and not of the kind $\beta_i(\ta(x),\ta(\uy), \ta(\uz),x, \uy,\uz)$).
This restriction has an important semantic effect, namely that \formulae~\eqref{eq:simpleflatdef} are equi-satisfiable to \formulae which are \emph{permutation-invariant}, in the following sense. The truth
value of an arithmetical formula or of a formula like $z=\sharp\{x\!\mid\! \alpha(\ta(x),\uy)\}$ is not affected by a permutation of the values of the $\ta(x)$ for $x\in [0, N)$, because 
$x$ does not occur free in $\alpha$ (permuting the values of the $\ta(x)$ may on the contrary change the value of a flat non simple sentence like 
$z=\sharp\{x\!\mid\! a(x)\leq x\}$). This `permutation invariance' will be exploited in the argument proving the correctness of decision procedure  of Theorem~\ref{thm:flatsimple} below. 
Formulae~\eqref{eq:simpleflatdef} 
%silvio
themselves 
are not permutation-invariant because of 
%the dependence on 
subterms $\ta(\uz), \ta(\uy)$, so we first show how to eliminate them up to satisfiability:

\begin{lemma}
 Simple E-flat \formulae are equi-satisfiable to disjunctions of per\-mu\-ta\-tion-invariant \formulae of the kind
  \begin{equation}\label{eq:simpleflatdef1}
   \exists \uz.~\alpha(\uy, \uz)\; \wedge \;\sharp\{x\mid \beta_1(\ta(x), \uy,\uz)\} =z_1 \wedge 
   \cdots\wedge \sharp\{x\mid \beta_K(\ta(x), \uy,\uz))\} =z_K~~~~~~
\end{equation}
\end{lemma}

\noindent
\textit{Proof.} Let us take a formula like~\eqref{eq:simpleflatdef}: we convert it to an equi-satisfiabòe disjunction of \formulae of the kind~\eqref{eq:simpleflatdef1}. The task is to eliminate terms $\ta(\uz)$, $\ta(\uy)$ by a series of guessings (each guessing will form the content
of a disjunct). Notice that
 we can apply the procedure of Lemma~\ref{lemma:nf} to eliminate the $\ta(\uz)$, but for the $\ta(\uy)$ we must operate differently (the method used  in Lemma~\ref{lemma:nf} introduced non simple abstraction terms). 
 
 Let us suppose that 
 $\uy:=y_1, \dots, y_m$ and that, after a first guess, $\alpha$ contains the conjunct $y_j< N$ for each
 $j=1, \dots, m$ (if it contains $y_j\geq N$,  we replace $a_s(y_j)$ by 0); after a second series of guesses, we can suppose also that
 $\alpha$ contains
 the conjuncts $y_{j_1}\neq y_{j_2}$ for $j_1\neq j_2$ (if it contains $y_{j_1}= y_{j_2}$, we replace $y_{j_1}$ by $y_{j_2}$ everywhere, making $y_{j_1}$ to disappear from the whole formula).
 In the next step, (i) we introduce for every $a\in \ta$ and for every $j=1, \dots, m$ a fresh variable $u_{aj}$, (ii) we replace everywhere $a(y_j)$ by 
 $u_{aj}$ and (iii) we conjoin to $\alpha$ the equalities $a(y_j)=u_{aj}$. In this way we get a formula of the following kind
 \begin{equation}\label{eq:tmp}
   \exists \uz.~\bigwedge_{a\in \ta, y_j\in \uy} a(y_j)=u_{aj} \wedge \alpha(\uy, \uu,\uz)\; \wedge \bigwedge_{l=1}^K\;\sharp\{x\mid \beta_l(\ta(x), \uy,\uu,\uz)\} =z_l 
\end{equation}
where $\uu$ is the tuple formed by the $u_{aj}$ (varying $a$ and $j$). We now make another series of guesses and conjoin to $\alpha$ either 
$u_{aj}=u_{a'j'}$ or $u_{aj}\neq u_{a'j'}$ for $(a,j)\neq (a', j')$. Whenever $u_{aj}=u_{a'j'}$ is conjoined, $u_{aj}$ is replaced by $u_{a'j'}$
everywhere, so that $u_{aj}$ disappears completely. The resulting formula still has the form~\eqref{eq:tmp}, but now the map $(a,j)\mapsto u_{aj}$ is not injective anymore (otherwise said,
 $u_{aj}$ now indicates the element from the tuple $\uu$ associated with the pair $(a,j)$ and we might have that the same $u_{aj}$ is associated 
 with different pairs $(a,j)$). 
 
 Starting from~\eqref{eq:tmp} so modified, let us define now the equivalence relation %$\approx$ 
 among the $y_j$ that holds
 between $y_j$ and $y_{j'}$
whenever for all $a\in \ta$ there is $u_a\in \uu$ such that $\alpha$ contains the equalities $a(y_j)=u_a$ and $a(y'_j)=u_a$. Each equivalence class 
$E$ is uniquely identified by the corresponding function $f_E$ from $\ta$ into $\uu$  (it is the function that for each $y_j\in E$ maps $a\in \ta$ to the $u_a\in \uu$ such that $\alpha$ contains 
as a conjunct the equality $a(y_j)=u_a$). Let $E_1, \dots, E_r$ be the equivalence classes and let $n_1, \dots, n_r$ be their cardinalities.
We claim that~\eqref{eq:tmp} is equisatisfiable to
\begin{equation}\label{eq:tmp1}
   \begin{aligned}
   \exists \uz.~ \alpha(\uy, \uu,\uz)\; 
   \wedge \bigwedge_{q=1}^r \;\sharp\{x\mid \bigwedge_{a\in \ta} a(x)=f_{E_q}(a)\}\geq n_q ~\wedge
   \\
   \wedge \bigwedge_{l=1}^K\;\sharp\{x\mid \beta_l(\ta(x), \uy,\uu,\uz)\} =z_l ~~~~~~~
   \end{aligned}
\end{equation}
In fact, satisfiability of~\eqref{eq:tmp} trivially implies the satisfiability of the formula~\eqref{eq:tmp1};
vice versa, since~\eqref{eq:tmp1} is permutation-invariant, if it is satisfiable we can modify any assignment satisfying it via a simultaneous permutation of the values of the $a\in \ta$ so as to produce  an assignment satisfying~\eqref{eq:tmp}.

We now need just the  trivial observation that the inequalities $\sharp\{x\mid \bigwedge_{a\in \ta} a(x)=f_{E_q}(a)\}\geq n_q $ can be 
replaced by the \formulae $\sharp\{x\mid \bigwedge_{a\in \ta} a(x)=f_{E_q}(a)\}= z'_q\;\wedge\;  z'_q\geq n_q $ (for fresh $z'_q$) in order
to match the syntactic shape of~\eqref{eq:simpleflatdef1}.
  $\hfill\dashv$

We can freely assume that quantifiers do not occur in simple flat  \formulae: this is without loss of generality because such \formulae are built up from 
arithmetic and basic \formulae.\footnote{ By the quantifier-elimination result for Presburger arithmetic, it is well-known that arithmetic \formulae 
are equivalent to quantifier-free ones. The same is true for basic \formulae  because they are obtained from arithmetic 
formulae by substitutions without capture. 
}

\begin{theorem}\label{thm:flatsimple}
 Satisfiability of simple flat  \formulae can be decided in NP (and thus it is an NP-complete problem).
\end{theorem}

\noindent
\textit{Proof.} 
 First, by applying the procedure of the 
 previous Lemma we can reduce to the problem of checking the satisfiability of \formulae of the kind
 \begin{equation}\label{eq:card00}
\alpha(\uy, \uz)~ \wedge~ \sharp\{x\mid \beta_1(\ta(x), \uy, \uz)\} =z_1~\wedge \cdots \wedge~ \sharp\{x\mid \beta_K(\ta(x),\uy, \uz )\} =z_K
\end{equation}
where $\alpha, \beta_1, \dots, \beta_K$ are basic
(notice also that each formula in the output of the procedure of the previous Lemma comes from a polynomial guess).

Suppose that $A_1(\ta(x), \uy, \uz), \dots, A_L(\ta(x), \uy, \uz)$ are the atoms occurring in 
$\beta_1, \dots, \beta_K$. For a Boolean assignment $\sigma$ to these atoms, we indicate with $\lbrack\!\lbrack \beta_j\rbrack\!\rbrack^\sigma$ the Boolean value (0 or 1) the 
 formula $\beta_l$ has under such assignment. We first claim that~\eqref{eq:card00} is satisfiable iff there exists \emph{a set of assignments} $\Sigma$ such that the formula
\begin{equation}\label{eq:cards}
\begin{aligned}
 &\alpha(\uy, \uz)~ \wedge \bigwedge_{\sigma\in \Sigma} \exists \uu \left(\bigwedge_{j=1}^L \epsilon_{\sigma(A_j)} A_j(\uu, \uy, \uz)\right)  \wedge
 \begin{aligned}
    \begin{bmatrix}
           z_{1} \\
           z_{2} \\
           \vdots \\
           z_{K}
     \end{bmatrix}
         = \sum_{\sigma\in \Sigma} v_{	\sigma} 
     \begin{bmatrix}
           \lbrack\!\lbrack \beta_1\rbrack\!\rbrack^\sigma \\
            \lbrack\!\lbrack \beta_2\rbrack\!\rbrack^\sigma \\
           \vdots \\
            \lbrack\!\lbrack \beta_K\rbrack\!\rbrack^\sigma
     \end{bmatrix}
     \wedge
   \end{aligned}   
   \\  
        &\wedge \sum_{\sigma\in \Sigma} v_{\sigma} = N      
        %&
        \wedge\bigwedge_{\sigma\in \Sigma} v_{\sigma} > 0
 \end{aligned}  
\end{equation}
is satisfiable (we introduced extra fresh variables $v_{\sigma}$, for $\sigma\in \Sigma$; notation $\epsilon_{\sigma(A_j)}$ is the same as in the proof of Lemma~\ref{lemma:nf}).
In fact, on one side, if~\eqref{eq:card00} is satisfiable under $V$, we can take as $\Sigma$ the set of assigments for which $\bigwedge_{j=1}^L \epsilon_{\sigma(A_j)} A_j(\ta(i), \uy, \uz)$
is true under $V$  for some $i\in [0, N)$ and for $v_{\sigma}$ the cardinality of the set of the $i\in [0, N)$ for which $\bigwedge_{j=1}^L \epsilon_{\sigma(A_j)} A_j(\ta(i), \uy, \uz)$ holds.
This choice makes~\eqref{eq:cards} true. 
Vice versa, if~\eqref{eq:cards} is true under $V$, in order to define the value of the tuple $\ta(i)$ (for $i\in [0,N)$), pick for every $\sigma\in \Sigma$ some $\uu_\sigma$ such that  
 $\bigwedge_{j=1}^L \epsilon_{\sigma(A_j)} A_j(\uu_{\sigma}, \uy, \uz)$ holds; then, supposing $\Sigma=\{\sigma_1, \dots, \sigma_h\}$, let $\ta(i)$ be equal to $\uu_{\sigma_1}$ for $i\in [0, v_{\sigma_1})$, to
  $\uu_{\sigma_2}$ for $i\in [v_{\sigma_1}, v_{\sigma_2})$, etc.
  Since we have that $ \sum_{\sigma\in \Sigma} v_{\sigma} = N $, the definition of the interpretation of the $\ta$ is complete (any other permutation of the values 
  $\ta(x)$ inside $[0, N)$ would fit as well).
  In this way, formula~\eqref{eq:card00} turns out to be true.

We so established that our original formula is satisfiable iff there is some $\Sigma$ such that~\eqref{eq:cards} is satisfiable;
%\footnote{
%Notice that, due to the presence of the variables $\uy, \uz$ which are constrained by the condition $\alpha(\uy, \uz)$, it is not possible to take as $\Sigma$ the set of all 
%assignments (in fact, there might not even be a maximum suitable set of assignments $\Sigma$).} 
the only problem we still have to face is that $\Sigma$ might be exponentially 
large. To reduce to a polynomial $\Sigma$, we use the same technique as in~\cite{cade21}. In fact, if~\eqref{eq:cards} is satisfiable, then the column vector $(z_1, \dots, z_K)^T$ is a linear combination with positive integer coefficients of the 0/1-vectors $(\lbrack\!\lbrack \beta_1\rbrack\!\rbrack^\sigma,
           \cdots,
            \lbrack\!\lbrack \beta_K\rbrack\!\rbrack^\sigma)^T$
and it is known from~\cite{ro} that, if this is the case, the same result can be achieved by assuming that at most $2K \log_2(4K)$ of the $v_{\sigma}$ are nonzero.  Thus polynomially many $\Sigma$ are sufficient and
for such $\Sigma$, a satisfying polynomial assignment for the existential Presburger formula~\eqref{eq:cards} is a polynomial certificate. 
 $\hfill\dashv$

{\subsection{ Some heuristics}\label{subsec:heuristics}
We discuss here some useful heuristics for  the satisfiability algorithm for simple flat \formulae (these heuristics have been implemented in our prototype).

1.- The satisfiability test involves all \formulae~\eqref{eq:cards} for each set of assignments $\Sigma$ having cardinality \emph{at most}   $M= \lceil 2K \log_2(4K)\rceil$ 
(actually, one can improve this bound, see~\cite{cade21}). If we replace in~\eqref{eq:cards}, for every $\sigma$, the conjunct 
$v_{\sigma} >0$ by $v_{\sigma} \geq 0$ and the conjunct
%$\bigwedge_{\sigma\in \Sigma} \exists \uu (\bigwedge_{j=1}^L \epsilon_{\sigma(A_j)} A_j(\uu, \uy, \uz))$
$\exists \uu\, (\bigwedge_{j=1}^L \epsilon_{\sigma(A_j)} A_j(\uu, \uy, \uz))$
by
%$\bigwedge_{\sigma\in \Sigma} \left(v_{\sigma}> 0 \to\exists \uu (\bigwedge_{j=1}^L \epsilon_{\sigma(A_j)} A_j(\uu, \uy, \uz))\right)$ 
$v_{\sigma}> 0 \to\exists \uu \,(\bigwedge_{j=1}^L \epsilon_{\sigma(A_j)} A_j(\uu, \uy, \uz))$, 
we can limit ourselves to the $\Sigma$ having cardinality \emph{equal to} $M$. This trick is useful if, for some reason, we prefer to go through any sufficient set of assignments (like  the set of all assignments supplied by some Boolean propagation, see below).

2.- There is no need to consider assignments $\sigma$ over the set of the atoms $A_j$ occurring in the $\beta_1,\dots, \beta_K$; any set of \formulae generating the $\beta_1, \dots, \beta_K$ by 
Boolean combinations fits our purposes. As a consequence, the choice of these
`atoms' is subject to case-by-case evaluations.

3.- Universally quantified \formulae of the kind $\forall x\,(0\leq x \wedge x< N \to \beta)$ can be turned into flat \formulae by rewriting them as
$N =\sharp\,\{ x\mid \beta\}$ (and in fact such universally quantified \formulae often occur in our benchmarks suite).
 These \formulae contribute to~\eqref{eq:card00} via the conjuncts of the kind
$z_i=N \wedge \sharp\{ x\mid \beta_i(\ta(x), \uy, \uz)\} = z_i$. It is quite useful to
consider the $\{\beta_{i_1}, \dots, \beta_{i_L}\}$ arising in this way as atoms (in the sense of point 2 above) and restrict to the assignments $\sigma$ 
such that $\sigma(\beta_{i_1})= \cdots = \sigma(\beta_{i_L})=1$. 

4.- Boolean propagation is a quite effective strategy to prune useless assignments; in our context, as soon as a partial assignment $\sigma$ is produced inside the assignments enumeration subroutine, 
an SMT solver is invoked to test the satisfiability of 
$\alpha(\uy, \uz)~ \wedge  \bigwedge_{j\in dom(\sigma)} \epsilon_{\sigma(A_j)} A_j(\uu, \uy, \uz)$; since this is a (skolemized) conjunct of~\eqref{eq:cards}, if 
the test is negative the current partial assignment is discarded and next partial assignment (obtained by complementing the value of the last assigned literal) is taken instead.

\input{experFASGEP}

\section{Conclusions, related and further work}\label{sec:conclusions}

We identified two  fragments of the rich syntax of Figure~\ref{fig:syntax} 
%(Presburger arithmetic with free function symbols and counting constraints) 
and we showed their decidability (for the second fragment we showed also a tight complexity bound). Since our fragments are closed under Boolean connectives, it is possible to use them not only in  bounded model checking
(where they can both give certifications and find bugs), but also
in order to decide whether an invariant holds or not. We implemented our algorithm for the weaker fragment and used it in some experiments.
As far as we know, this is the first implementation of a \emph{complete} algorithm for a fragment of arithmetic with arrays and counting capabilities for interpreted sets.   
 
 Since one of the major intended applications concerns fault-tolerant distributed systems, we  briefly review and compare here some recent work in the area.
Papers~\cite{konnov_cav},~\cite{konnov_concur},~\cite{konnov_fmcad} represent a very interesting and effective research line, where cardinality constraints are not directly handled but abstracted away 
using interval abstract domains and counters. As a result, a remarkable amount of algorithms are certified, although the method might suffer of some lack of expressiveness for more complex examples.

On the contrary, paper~\cite{sharpie} directly handles cardinality constraints for interpreted sets; nontrivial  invariant properties are synthesized and checked, 
based on Horn constraint solving technology. At the level of decision procedures, some incomplete inference schemata are employed (completeness is nevertheless showed for 
array updates against difference bounds constraints).

Paper~\cite{consensus} introduces a very expressive logic, specifically tailored to handle consensus problems (whence the name `consensus logic' $CL$). Such logic employs arrays with values into power set types, hence it is
situated in a higher order logic context. Despite this, our flat fragment is not fully included into $CL$, because we allow arithmetic constraints on the sort of 
indexes  and also mixed constraints between indexes and data: in fact, we have a unique sort for indexes and data, leading to the possibility of writing typically non permutation-invariant \formulae like $\sharp \,\{x \mid a(x) + x = N\} = z$. As pointed out in~\cite{AlbertiGS14}, this mono-sorted approach is useful in the analysis of programs, when pointers to the memory (modeled
as an array) are stored into array variables. From the point of view of deduction, the paper~\cite{consensus} uses an incomplete algorithm in order to certify   invariants. A smaller decidable fragment (identified via several syntactic restrictions) is introduced in the final part of the paper; 
the sketch of the decidability proof supplied for this smaller fragment uses bounds for minimal solutions of Presburger \formulae as well as Venn regions decompositions in order to build models 
where all  nodes in the same Venn region share the same value for their function symbols.

In future, we plan to extend both our tool \textsc{ArCa} and our results in order  to deal with more complex verification problems. Although it won't be easy to find richer fragments inheriting all the nice properties we discovered in this paper, we are confident that concrete applications will suggest viable effective extensions.

%\newpage
\bibliographystyle{plain}
\bibliography{FASGEP_2016}

\vfill\eject 
\appendix

\input{appendix_FASGEP}

\end{document}

%% file: packages.tex
\usepackage{amsmath}
\usepackage{amstext}
\usepackage{amssymb}
\usepackage{amsfonts}
\usepackage{xspace}
\usepackage{graphicx}
\usepackage{url}
\usepackage{subfigure}
\usepackage{mdwlist}
\usepackage{courier}
\usepackage{lscape}
\usepackage{comment}
\usepackage{wrapfig}
\usepackage{multirow}
\usepackage{bussproofs}
\usepackage{pdftricks}
\usepackage{subfigure}
\begin{psinputs}
\usepackage{pstricks}
\usepackage{color}
\usepackage{pstcol}
\usepackage{pst-plot}
\usepackage{pst-tree}
\usepackage{pst-eps}
\usepackage{multido}
\usepackage{pst-node}
\usepackage{pst-eps}
\end{psinputs}
\usepackage{cite}
\makeatletter
\newif\if@restonecol
\makeatother

\usepackage{footmisc}
\usepackage{balance}
\usepackage{enumerate}
\usepackage[normalem]{ulem}
\usepackage{hyperref}
\usepackage{alltt}

\usepackage{epsfig}
\usepackage{manfnt}
\usepackage{graphicx}
\usepackage{color}
\usepackage{multirow}
\usepackage{tabularx,colortbl}
\usepackage{alltt}
\usepackage{bussproofs}
\usepackage{algorithm}
\usepackage{algorithmicx}
\usepackage{pifont}
\usepackage{cancel}
\usepackage{rotating}

\usepackage{float,lipsum}
\floatstyle{boxed}
\restylefloat{table}
\restylefloat{figure}

%% file: macro.tex
\newcommand{\formulae}{formul\ae\xspace}

\newcommand{\dpll}[1]{{\sc DPLL}\xspace}

\newcommand{\COMMENT}[1]{}

\newcommand{\ux}{\ensuremath{\underline x}}
\newcommand{\uu}{\ensuremath{\underline u}}

\newcommand{\uy}{\ensuremath{\underline y}}
\newcommand{\uz}{\ensuremath{\underline z}}

\newcommand{\up}{\ensuremath{\underline p}}

\renewcommand{\int}{\ensuremath {\mathcal I}}

\newcommand{\ta}{\ensuremath{{\bf a}}\xspace}

%% file: experFASGEP.tex
\section{Examples and experiments}\label{sec:experiments}

We implemented a prototype \textsc{ArCa-Sat}\footnote{\textsc{ArCa} stands for \emph{Array with Cardinalities}.} producing out of  simple E-flat \formulae~\eqref{eq:card00} the  proof obbligations~\eqref{eq:cards} (written  in SMT-LIB2 format), exploiting the heuristics 
explained in Section~\ref{subsec:heuristics}. To experiment the feasibility of our approach for concrete verification problems, we also implemented a (beta) version of a tool called \textsc{ArCa} producing out of the
specification of a parametric distributed system and of a safety-like problem, some E-flat simple \formulae whose unsatisfiability formalizes invariant-checking and bounded-model checking problems. A script executing in sequence 
\textsc{ArCa},  \textsc{ArCa-Sat} and \textsc{z3} can then solve such problems by reporting a `sat/unsat' answer.

A system is specified via a pair of flat (simple) \formulae $\iota(\up)$ and $\tau(\up, \up')$
%silvio mettere una formula \tau sola o più d'una non cambia perchè si fa la disgiuzione (questa cosa è abbastanza standard da poterla lasciare implicita)
and a safety problem via a further formula $\upsilon(\up)$ (here the $\up$ are parameters and array-ids, the $\up'$ are renamed copies of the $\up$).
A bounded model checking problem is the problem of checking whether the formula
$$
\iota(\up_0) \wedge \tau(\up_0, \up_1)\wedge\cdots\wedge \tau(\up_n, \up_{n+1})\wedge \upsilon(\up_{n+1})
$$
is satisfiable for a fixed $n$. An invariant-cheking problem, given also a formula $\phi(\up)$, is the problem of checking whether the three \formulae
$$
\iota(\up)\wedge \neg \phi(\up), \quad \phi(\up)\wedge \tau(\up, \up') \wedge \neg \phi(\up'), \quad \phi(\up)\wedge \upsilon(\up)
$$
are unsatisfiable. Notice that since all our algorthms terminate and are sound and complete, the above  problems are always solved by the above tool combination (if enough
computation resources are available). Thus, our technique is able \emph{both to make safety certifications and to find bugs}.

To validate our technique, in the following we describe in detail the formalization of the send-receive broadcast primitive (SRBP) in \cite{Srikanth87}.  SRBP  is used as a basis to synchronize clocks in systems where  processes may fail in sending and/or receiving messages.  Periodically, processes broadcast the virtual time to be adopted by all, as a \textit{(session s)} message.  Processes that accept this message set $s$ as their current time.  SRBP aims at guaranteeing the following properties:
\begin{description}
\item [Correctness:] if at least $f+1$ correct processes broadcast the message \textit{(session s)}, all correct processes accept the message.
\item [Unforgeability:] if no correct process broadcasts \textit{(session s)}, no correct process accepts the message.
\item [Relay:] if a correct process accepts \textit{(session s)}, all correct processes accept it.
\end{description}
where $f < N/2$ is the number of processes failing during an algorithm run, with $N$ the number of processes in the system.  Algorithm \ref{SRBPcode} shows the pseudo-code.

%%%%%%%
\begin{algorithm}[t]
%\begin{center}
%\begin{tabular}{|c|}\hline
%\begin{minipage}{.95\textwidth}
\begin{footnotesize}
\begin{tabbing}
aaaa\=bb\=cccc\=dddd\=eeee\=ffff   \kill
{\bf Initialization:}  \\
\> To broadcast a \textit{(session s)} message, a correct process sends \textit{(init, session s)} to all.  \\
{\bf{End Initialization}}  \\
{\bf for} each correct process:  \\
1.\> \textbf{if} received \textit{(init, session s)} from at least $f+1$ distinct processes \textbf{or}  \\
2.\> \> received  \textit{(echo, session s)} from any process \textbf{then}  \\
3.\> \> \> accept  \textit{(session s)};  \\
4.\> \> \> send  \textit{(echo, session s)} to all;  \\
5.\> \textbf{endif}  \\
{\bf end for}
\end{tabbing}
\end{footnotesize}
%\end{minipage} \\ \hline
%\end{tabular}
%\end{center}
\caption{\label{SRBPcode}Pseudo-code for the send-receive broadcast primitive.}
\end{algorithm}
%%%%%%%

We model SRBP as follows: $IT(x)$ is the initial state of a process $x$; it is $s$ when $x$ broadcasts a \textit{(init, session s)} message, and 0 otherwise.  $SE(x)=s$ indicates that $x$ has broadcast its own echo. $AC(x)=s$ indicates that $x$ has accepted \textit{(session s)}.  Let $pc$ be the program counter, $r$ the round number, and $G$ a flag indicating whether one round has been executed.  We indicate with $F(x)=1$ the fact that $x$ is faulty, and $F(x)=0$ otherwise.  Finally, $CI(x)$ and $CE(x)$ are the number of respectively inits and echoes received.  
In the following, $\forall x$ means $\forall x \in [0, N)$.  
%In this section, we use $N-1$ where in the rest of the paper we used $N$ (thus for instance $\forall x$ means $\forall x \in [0, N-1]$, etc.).
Some sentences are conjoined to all our proof obligations, namely:
$ \#\{x | F(x)=0\} + \#\{x | F(x) = 1\} = N \wedge \#\{x | F(x) = 1\} < N/2$.
For the Correctness property, we write $\iota_c$ as follows:
%%%
{\footnotesize
\begin{eqnarray}
\iota_c & := & pc=1 \wedge r=0 \wedge G=0 \wedge s \not =0 \wedge \\
& & \#\{x | IT(x)=0\} + \#\{x | IT(x) = s\} = N \wedge \\
& & \#\{x | F(x) = 0 \wedge IT(x)=s\} \geq (\#\{x | F(x) = 1\} +1) \wedge \\
& & \forall x. SE(x) = 0 \wedge AC(x) = 0 \wedge CI(x) = 0 \wedge CE(x) = 0
\end{eqnarray}
}
%%%
where we impose that the number of correct processes broadcasting the init message is at least the number of faulty processes, $f$, plus 1.  It is worth to notice that -- from the above definition -- our tool produces a specification that is checked for any $N \in \mathbb{N}$ number of processes.  The constraints on $IT$ allow to verify all admissible assignments of 0 or $s$ to the variables.  Similarly for $F(x)$.

The algorithm safety is verified by checking that
%whether 
the bad properties 
%negations 
cannot be reached from the initial state.  For Correctness, we set
$\upsilon_c := pc=1 \wedge G=1 \wedge \#\{x | F(x) =0 \wedge AC(x)=0\} >0$,
that is, Correctness is not satisfied if -- after one round -- some correct process exists that has yet to accept.
The algorithm evolution is described by two transitions: $\tau_1$ and $\tau_2$.  The former allows to choose the number of both inits and echoes received by each process.  The latter describes the actions in Algorithm \ref{SRBPcode}.

{\footnotesize
%%%
\begin{eqnarray*}
\tau_1 & := & pc=1 \wedge pc'=2 \wedge r'=r \wedge G' =G \wedge s'=s \wedge \exists K1, K2, K3, K4. \\
& & K1 = \#\{x | F(x) =0 \wedge IT(x)=s\} \wedge K2 = \#\{x | F(x) = 0 \wedge SE(x)=s\} \wedge \\
& & K3 = \#\{x | F(x) = 1 \wedge IT(x)=s\} \wedge K4 = \#\{x | F(x) = 1 \wedge SE(x)=s\} \wedge \\
& & \forall x. F(x) =0 \Rightarrow (CI'(x) \geq K1 \wedge CI'(x) \leq (K1+K3) \wedge CE'(x) \geq K2 \wedge \\ 
& & CE'(x) \leq (K2+K4)) \wedge \\
& & \forall x. F(x) =1 \Rightarrow (CI'(x) \geq 0 \wedge CI'(x) \leq (K1+K3) \wedge CE'(x) \geq 0 \wedge \\ 
& & CE'(x) \leq (K2+K4)) \wedge \\
& & \forall x. IT'(x) = IT(x) \wedge SE'(x) = SE(x) \wedge AC'(x) = AC(x) \\
\tau_2 & := & pc=2 \wedge pc' = 1 \wedge r' = (r+1) \wedge s'=s \wedge G'=1 \wedge \\
& & \forall x. (CI(x) \geq \#\{x | F(x)=1\}+1 \Rightarrow SE'(x)=s \wedge AC'(x)=s ) \wedge \\
& & \forall x. (CI(x) < \#\{x | F(x)=1\}+1 \wedge CE(x) \geq 1 \Rightarrow SE'(x)=s \wedge AC'(x)=s) \wedge \\
& & \forall x.  (CI(x) < \#\{x | F(x)=1\}+1 \wedge CE(x) < 1 \Rightarrow SE'(x)=0 \wedge AC'(x)=0) \wedge \\
& & \forall x. IT'(x) = IT(x) \wedge CI'(x) = CI(x) \wedge CE'(x) = CE(x)
\end{eqnarray*}
%%%
}

The same two transitions are used to verify both the Unforgeability and the Relay properties, for which however we have to change the initial and final formula.  For Unforgeability, (13) in $\iota$ changes as $... \wedge \#\{x | F(x) = 0 \wedge IT(x)=0\} = \#\{x | F(x) = 0\} \wedge ...$; while $\upsilon_u := pc=1 \wedge G=1 \wedge \#\{x | F(x) =0 \wedge AC(x)=s\} >0$.
In $\iota_u$ we say that all non-faulty processes have $IT(x)=0$.  Unforgeability is not satisfied if some correct process accepts.  For Relay, we use:
%%%
{\footnotesize
\begin{eqnarray*}
\iota_r & := & pc=1 \wedge r=0 \wedge s \not =0 \wedge G=0 \wedge \\
& & \#\{x | F(x) = 0 \wedge AC(x) = s \wedge SE(x)=s\} = 1 \wedge \\
& & \#\{x | AC(x)=0 \wedge SE(x)=0\} = (N-1) \wedge \#\{x | AC(x)=s \wedge SE(x)=s\} = 1 \wedge \\
& & \forall x. IT(x) = 0 \wedge CI(x) = 0 \wedge CE(x) = 0
\end{eqnarray*}
}
%%%
while $\upsilon_r = \upsilon_c$.  In this case, we start the system in the worst condition: by the hypothesis, we just know that one correct process has accepted.  Upon acceptance, by the pseudo-code, it must have sent an echo.  All the other processes are initialized in an idle state.  We also produce an unsafe model of Correctness: we modify $\iota_c$ by imposing that just $f$ correct processes broadcast the init message.

%%%%%
\begin{table}
\caption{\label{tabellone}Evaluated algorithms and experimental results.}
\centering
{\scriptsize
\begin{tabular}{|l|c|c|c|c|c|} \hline
\textbf{Algorithm} & \textbf{Property} & \textbf{Condition} & \textbf{Problem} & \textbf{Outcome} & \textbf{Time (s.)} \\ \hline
SRBP \cite{Srikanth87} & Correctness & $\geq (f+1)$ init's & \textsc{bmc} & safe & 0.82   \\ \hline
SRBP \cite{Srikanth87} & Correctness & $\leq f$ init's &  \textsc{bmc} & unsafe & 2.21  \\ \hline
SRBP \cite{Srikanth87} & Unforgeability & $\geq (f+1)$ init's & \textsc{bmc} & safe & 0.85  \\ \hline
SRBP \cite{Srikanth87} & Relay & $\geq (f+1)$ init's & \textsc{bmc} & safe & 1.93   \\ \hline
BBP \cite{toueg87} & Correctness & $N > 3f$ & \textsc{bmc} & safe & 6.17  \\ \hline
BBP \cite{toueg87} & Unforgeability & $N > 3f$ & \textsc{bmc} & safe & 0.25  \\ \hline
BBP \cite{toueg87} & Unforgeability & $N \geq 3f$ & \textsc{bmc} & unsafe & 0.25 \\ \hline
BBP \cite{toueg87} & Relay & $N > 3f$ & \textsc{bmc} & safe & 1.01  \\ \hline
OT \cite{sharpie} & Agreement & threshold $> 2N/3$ & \textsc{ic} & safe & 4.20 \\ \hline
OT \cite{sharpie} & Agreement & threshold $> 2N/3$ & \textsc{bmc} & safe & 278.95  \\ \hline
OT \cite{sharpie} & Agreement & threshold $\leq 2N/3$ & \textsc{bmc} & unsafe & 17.75  \\ \hline
OT \cite{sharpie} & Irrevocability & threshold $> 2N/3$ & \textsc{bmc} & safe & 8.72  \\ \hline
OT \cite{sharpie} & Irrevocability & threshold $\leq 2N/3$ & \textsc{bmc} & unsafe & 9.51  \\ \hline
OT \cite{sharpie} & Weak Validity & threshold $> 2N/3$ & \textsc{bmc} & safe & 0.45  \\ \hline
OT \cite{sharpie} & Weak Validity & threshold $\leq 2N/3$ & \textsc{bmc} & unsafe & 0.59  \\ \hline
UV \cite{heardof} & Agreement & ${\cal P}_{nosplit}$ violated & \textsc{bmc} & unsafe & 4.18  \\ \hline
UV \cite{heardof} & Irrevocability & ${\cal P}_{nosplit}$ violated & \textsc{bmc} & unsafe & 2.04  \\ \hline
UV \cite{heardof} & Integrity & - & \textsc{bmc} & safe & 1.02 \\ \hline
U$_{T,E,\alpha}$  \cite{Biely07} & Integrity & $\alpha=0 \wedge {\cal P}_{safe}$ & \textsc{bmc} & safe & 1.16  \\ \hline
U$_{T,E,\alpha}$  \cite{Biely07} & Integrity & $\alpha=0 \wedge \neg{\cal P}_{safe}$ & \textsc{bmc} & unsafe & 0.83  \\ \hline
U$_{T,E,\alpha}$  \cite{Biely07} & Integrity & $\alpha=1 \wedge {\cal P}_{safe}$ & \textsc{bmc} & safe & 5.20  \\ \hline
U$_{T,E,\alpha}$  \cite{Biely07} & Integrity & $\alpha=1 \wedge \neg{\cal P}_{safe}$ & \textsc{bmc} & unsafe & 4.93  \\ \hline
U$_{T,E,\alpha}$  \cite{Biely07} & Agreement & $\alpha=0 \wedge {\cal P}_{safe}$ & \textsc{bmc} & safe & 59.80  \\ \hline
U$_{T,E,\alpha}$  \cite{Biely07} & Agreement & $\alpha=0 \wedge \neg{\cal P}_{safe}$ & \textsc{bmc} & unsafe & 7.78  \\ \hline
U$_{T,E,\alpha}$  \cite{Biely07} & Agreement & $\alpha=1 \wedge {\cal P}_{safe}$ & \textsc{bmc} & safe & 179.67  \\ \hline
U$_{T,E,\alpha}$  \cite{Biely07} & Agreement & $\alpha=1 \wedge \neg{\cal P}_{safe}$ & \textsc{bmc} & unsafe & 31.94  \\ \hline
%
%%%MESI \cite{Patel84} & & & \textsc{bmc} & safe &  \\ \hline
MESI \cite{Patel84} & cache coherence & - & \textsc{ic} & safe & 0.11  \\ \hline
%%%MOESI \cite{MOESI} & & & \textsc{bmc} & safe &  \\ \hline
MOESI \cite{MOESI} & cache coherence & - & \textsc{ic} & safe & 0.08  \\ \hline
%%%Dekker \cite{Dekker} & & & \textsc{bmc} & safe &  \\ \hline
Dekker \cite{Dekker} & mutual exclusion & - & \textsc{ic} & safe & 2.05  \\ \hline
\end{tabular}
}
\end{table}
%%%%%
In Table \ref{tabellone}, we report the results of validating these and other models with our tool.  In the first column, the considered algorithm is indicated.  The second column indicates the property to be verified; the third column reports the conditions of verification.  In the fourth column, we indicate whether we consider either a bounded model checking (\textsc{bmc}) or an invariant-checking (\textsc{ic}) problem.  The fifth column supplies the obtained results.  The sixth column shows the time jointly spent by \textsc{ArCa},  \textsc{ArCa-Sat} and \textsc{z3} for the verification, considering for \textsc{bmc} the sum of the times spent for every traces of length up to 10.  We used a PC equipped with Intel Core i7  processor and operating system Linux Ubuntu 14.04 64 bits.
We focused on \textsc{bmc} problems as they produce longer formulas thus stressing more the tools.
Specifically, following the example above, we modeled: 
\begin{itemize}
\item the byzantine broadcast primitive (BBP) \cite{toueg87} used to simulate authenticated broadcast in the presence of malicious failures of the processes, 
\item the one-third algorithm (OT) \cite{sharpie} for consensus in the presence of benign transmission failures, 
\item the Uniform Voting (UV) algorithm \cite{heardof} for consensus in the presence of benign transmission failures, 
\item the U$_{T,E,\alpha}$ algorithm \cite{Biely07} for consensus in the presence of malicious transmission failures, 
\item the MESI \cite{Patel84} and MOESI \cite{MOESI} algorithms for cache coherence,
\item the Dekker's algorithm \cite{Dekker} for mutual exclusion. 
\end{itemize}
All the models, together with our tools to verify them, are available at \url{http://users.mat.unimi.it/users/ghilardi/arca}.

As far as the processing times are concerned, we observed that on average z3 accounts for around 68$\%$ of the processing time, while \textsc{ArCa} and \textsc{ArCa-Sat} together account for the remaining 32$\%$.
Indeed, the SMT tests performed by \textsc{ArCa-Sat} are lightweight -- as they only prune assignments -- yet effective, as they succeed in reducing the number of assignments of at least one order of magnitude.

%% file: appendix_FASGEP.tex
\section{Counting constraints in Presburger arithmetic}\label{app:counting}

We report here a proof of Theorem~\ref{thm:NS}. This is not an original result and we will not try to optimize it, rather we just rewrite  proofs inside our notations, 
trying at the same time to supply the reader some intuitive evidence about the reasons why the theorem
holds.

Take a constraint formula  $\phi$ (this is a formula built up from the  grammar of Figure~\ref{fig:syntax} without using
 array-ids). For every atom $A$ occurring in it (i.e. for every subformula of the kind $t_1< t_2, t_1= t_2$ ot $t_1\equiv_n t_2$)
 and for every outermost occurrence of a subterm of the kind $\sharp\, \{ x\mid \psi\}$ in $A$, pick a fresh variable $z$ and replace $A$
 in $\phi$ with 
 $\exists z\, (z= \sharp\, \{ x\mid \psi\} \wedge A')$, where $A'$ is obtained from $A$ replacing the occurrence of the 
 subterm $\sharp\, \{ x\mid \psi\}$ by $z$. If we call $\phi'$ the resulting formula, it is clear that $\phi$ and $\phi'$ are equivalent.
 
 By repeating this procedure, we can transform any constraint formula (up to equivalence) into a constraint formula built up according to the following
 more restricted instructions:
 %of Figure~\ref{fig:syntax1}.
 \begin{description}
  \item[{\rm (i)}] \emph{arithmetic terms} are built up from numerals $0, 1, \dots $, individual variables $x, y, z, \dots $ and parameters $ M, N, \dots $ using $+$ and $-$;
  \item[{\rm (ii)}] \emph{arithmetic atoms} are expressions of the kind $t_1<t_2, t_1=t_2, t_1\equiv_n t_2$, where $t_1, t_2$ are arithmetic terms;
  \item[{\rm (iii)}] \emph{arithmetic \formulae} are built up from arithmetic atoms using $\wedge, \neg, \exists$ (actually, $\exists$ is redundant, given that quantifier-elimination holds);
  \item[{\rm (iv)}] \emph{constraint atoms} are either arithmetic atoms or expressions of the form $y=\sharp\{ x\mid \alpha\}$, where $\alpha$ is an arithmetic formula;
  \item[{\rm (v)}] \emph{constraint \formulae} are built up from constraint atoms using  $\wedge, \neg, \exists$.
 \end{description}

Recall that we interpret  $\sharp\{ x\mid \alpha\}$ as the cardinality of the set formed by the $x$ such that $0\leq x <N $ and $\alpha(x)$ is true. Thus, if we want to translate our 
constraint atoms into the terminology of~\cite{schweikhart}, we must translate $y=\sharp\{ x\mid \alpha\}$ as $\exists^{=y} x \, (0\leq x \wedge x< N \wedge \alpha)$
(in this sense, our formalism apparently looks slightly less expressive and the procedure below has few less cases than~\cite{schweikhart}).

 It is then evident that Theorem~\ref{thm:NS} is proved once 
we show the following 
\begin{theorem}\label{thm:NS1}
 Every constraint atom is equivalent to an arithmetic formula.
\end{theorem}

\noindent
\textit{Proof.}
The following \emph{special case} of Theorem~\ref{thm:NS1} is easy: if $x$ does not occur in the arithmetic terms $t_1, t_2, t_3$, then 
the constraint atom  
\begin{equation}\label{eq:basicleft}
 y= \sharp\, \{ x\mid t_1\leq x \wedge x < t_2 \wedge x\equiv_n t_3\}
\end{equation}
 is equivalent to the formula
 \begin{equation}\label{eq:basicright}
  \begin{aligned}
  &
  \exists z\;\left(
      \begin{aligned}
       & t_1\leq z \;\wedge\; z < t_2 \;\wedge\; z\equiv_n t_3 \;\wedge\; 
       \\ &  \forall z' (t_1\leq z \;\wedge\; z < t_2\; \wedge\; z\equiv_n t_3\;\to\; z\leq z') ~\wedge
       \\ & y =\lceil {t_2-z\over n}\rceil
      \end{aligned}
    \right) \vee 
  \\
  &
  \vee
  \left(
    \begin{aligned}
    & \neg \exists z \;(t_1\leq z \;\wedge\; z < t_2\; \wedge z\equiv_n t_3 ) ~\wedge 
    \\ & y=0 
    \end{aligned}
     \right)
  \end{aligned}
\end{equation}
What formula~\eqref{eq:basicright} says is that either there is no $z\in [t_1, t_2)$ such that $z\equiv_n t_3$ (and then $y=0$) or there is such a $z$ (and then, taking the minimum such $z$,
we have that $y=\lceil {t_2-z\over n}\rceil$). Notice that the condition  $y=\lceil {t_2-z\over n}\rceil$ can be expressed in Presburger arithmetic via $\bigvee_{l=0}^{n-1}  (ny=l+t_2-z)$.\footnote{
We use obvious abbreviations like $ny=y+\cdots + y$ ($n$-times).
}

We now show how to reduce to the above special case, using the series of Lemmas of Subsection~\ref{subsec:lemmmas} below.

Consider in fact a constraint atom $y=\sharp\{ x\mid \alpha\}$; we can suppose that $\alpha$ is quantifier-free because Presburger arithmetic enjoys quantifier elimination. We can also eliminate negations using the equivalences  $t\neq u\leftrightarrow (t<u\vee u<t)$, and $t\not < u\leftrightarrow u\leq t$,\footnote{
Here $u\leq t$ stands for $u=t\vee u<t$.} 
and $t\not\equiv_n u\leftrightarrow \bigvee_{l=1}^{n-1} (t\equiv_n u+l)$.
Using Lemma~\ref{lem:pairwise} and  disjunctive normal forms arising from Venn's regions analysis, we can freely assume that $\alpha$ is a conjunction
of arithmetic atoms; atoms in which $x$ does not occur can be eliminated using Lemma~\ref{lem:independent}.
By normalizing terms as linear polynomials, we can further limit to atoms of the kinds
$$
kx=t, ~~t< kx,~~ kx<t,  ~~kx\equiv_n t,
$$
where  $k,n \geq 1$ and where $t$ is an arithmetic term in which $y$ does not occur. 
By Lemma~\ref{lem:unique}, we can solve the case where there are atoms of the kind $kx=t$. If there are no atoms like that,
using Lemmas~\ref{lem:ineq2},\ref{lem:ineq1},\ref{lem:c},\ref{lem:congruence}, we can freely assume that $k=1$.\footnote{ In case an inconsistent condition
arises according to Lemma~\ref{lem:congruence}(i), the constraint atom is replaced by $z=0$.}

To sum up, we are left with a constraint atom $y=\sharp\{ x\mid \alpha\}$ where $\alpha$ is of the kind
$$
\bigwedge_{i=1}^q t_i \leq x ~\wedge~\bigwedge_{j=1}^r x<u_j~ \wedge ~\bigwedge_{h=1}^s x\equiv_{n_s}v_h~~.
$$
(we  used  obvious equivalences like $t<x \leftrightarrow t+1\leq x$).
We can now reduce to $q=1$ and $r=1$ by making a disjunctive guess for determining the biggest $t_i$ and the lowest $u_j$ 
(Lemma~\ref{lem:independent} is then used to eliminate atoms where $x$ does not occur).\footnote{We  assume that $r,q\geq 1$
because $0\leq x$ and $x<N$ must be included among the conjuncts of $\alpha$.}
By Lemma~\ref{lem:lcm} and Lemmas~\ref{lem:pairwise},\ref{lem:independent}, we can also freely assume that $s=1$. Thus we finally end up in the special case above.
 $\hfill\dashv$

\subsection{Ingredient Lemmas}\label{subsec:lemmmas}

We collect here the facts we used  in the above proof (they are all almost obvious).

\begin{lemma}\label{lem:pairwise} If the formulae $\alpha_i$ are pairwise inconsistent, then 
 $y=\sharp\{ x\mid \bigvee_{i=1}^n\alpha_i\}$ is equivalent to
 $$
 \exists z_1\cdots \exists z_n \;(\bigwedge_{i=1}^n z_i=\sharp\{ x\mid \alpha_i\} \wedge y= \sum_{i=1}^n z_i)~~.
 $$
\end{lemma}

\begin{lemma}\label{lem:independent}
 If $x$ does not occur in $\beta$, then $y=\sharp\{ x\mid \alpha\;\wedge\; \beta\}$ is equivalent to
 $$
 (\neg \beta \wedge y=0)\;\vee\; (\beta \;\wedge\; y=\sharp\{ x\mid \alpha\})~~.
 $$
\end{lemma}

\begin{lemma}\label{lem:unique}
 If $x$ does not occur in $t$, then $y=\sharp\{ x\mid \alpha\;\wedge\; kx=t\}$ is equivalent to
 $$
 (y=1 \wedge \exists x\;(\alpha\;\wedge\; kx=t)) \vee (y=0 \wedge\neg \exists x\;(\alpha\;\wedge\; kx=t))~~.
 $$
\end{lemma}

\begin{lemma}\label{lem:c}
 Let $t$ be an arithmetic term where $x$ does not occur; then the constraint atom  $y=\sharp\{ x\mid \alpha\;\wedge\; t\equiv_n kx\}$ is equivalent to
 $$
 \bigvee_{l=0}^{n-1} (t\equiv_n l \wedge y=\sharp\{ x\mid \alpha\;\wedge\; l\equiv_n kx\})
 $$
\end{lemma}

Next two lemmas just report basic arithmetic facts:

\begin{lemma}\label{lem:congruence}
 For $n, l\geq 1$ and $k\geq 0$, let $g:=gcd(l,n)$; consider the linear congruence  $lx\equiv_n k$; we have that
 \begin{description}
  \item[{\rm (i)}] if $g\mid k$ does not hold, then $lx\equiv_n k$ is inconsistent (i.e. it does not have a solution);
  \item[{\rm (ii)}] if $g\mid k$ holds, then one can compute $n', k'$ such that
 $lx\equiv_n k$ is equivalent to $x\equiv_{n'} k'$.
 \end{description}
\end{lemma}

\noindent
\textit{Proof.}
Item (i) is obvious, because, if 
$lx\equiv_n k$ has a solution, then we have $lx-qn=k$ for some $q$. Suppose now that  $g\mid k$ holds and
 let $l':=l/g$, $n':=n/g$, $\tilde k:=k/g$. Since $gcd(n',l')=1$ and since gcd's can be expressed as linear combinations, there exists $l''$ such that $l'l''\equiv_{n'} 1$.
  But then $lx\equiv_n k$  is the same as 
 $l'gx\equiv_{gn'} \tilde k g$ which is equivalent to $l'x\equiv_{n'} \tilde k$, i.e. to $x\equiv_{n'} k'$, for $k':=l''\tilde k$.
 $\hfill\dashv$

\begin{lemma}\label{lem:lcm}
Let $k_1, \dots, k_m\in \mathbb Z$, $n_1, \dots, n_m\geq 1$ and $l:=lcm(k_1, \dots, k_m)$; then
$x\equiv_{n_1} k_1\;\wedge \cdots \wedge\; x\equiv_{n_m} k_m$ is equivalent to
$$
\bigvee_{r=0}^{l-1} ( x\equiv_l r \;\wedge\;  r\equiv_{n_1} k_1\;\wedge \cdots \wedge\; r\equiv_{n_m} k_m)~~.
$$
\end{lemma}

\begin{lemma}\label{lem:impl}
 Let $t$ be an arithmetic term, $n\geq 1, q\in \mathbb Z$ and $l\in\{0, \dots, n-1\}$; the following implications are valid
 \begin{equation*}
  \begin{aligned}
   & t-1= nq+l & \to & ~~\forall z\,(nz< t \leftrightarrow z< q\!+\!1)~~~~ \\
   & t+1= nq+l & \to & ~~\forall z\,(t<nz \leftrightarrow q\!-\!1<z) \\
  \end{aligned}
 \end{equation*}
\end{lemma}

\noindent
\textit{Proof.} We prove the validity of the first implication (the second is shown in an analogous way). Assume $t-1= nq+l$; then $nz< t$ is equivalent to
$n(z-q)\leq l$. This is the same as $z-q\leq 0$ (i.e. to $z< q+1$, as wanted), because otherwise we have $z-q\geq 1$ which implies $l\geq n(z-q)\geq n$, absurd.
  $\hfill\dashv$

\begin{lemma}\label{lem:ineq1}
 Let $t$ be an arithmetic term where $x$ does not occur; then the constraint atom  $y=\sharp\{ x\mid \alpha\;\wedge\; nx < t\}$ is equivalent to
 $$
 \bigvee_{l=0}^{n-1}  \exists q\, (t-1= nq+l \;\wedge\; y=\sharp\{ x\mid \alpha\;\wedge\; x< q+1\})~~.
 $$
\end{lemma}

\noindent
\textit{Proof.}  By the existence of quotients and remainders,  $y=\sharp\{ x\mid \alpha\;\wedge\; nx\leq t\}$ is equivalent to
 $\bigvee_{l=0}^{n-1}  \exists q\, (t-1= nq+l) \; \wedge\; y=\sharp\{ x\mid \alpha\;\wedge\; nx< t\}$, i.e. to
$$
 \bigvee_{l=0}^{n-1}  \exists q\, (t-1= nq+l \;\wedge\; y=\sharp\{ x\mid \alpha\;\wedge\; nx< t\})~~.
 $$
Now it is sufficient to apply the previous lemma.
 $\hfill\dashv$

\begin{lemma}\label{lem:ineq2}
 Let $t$ be an arithmetic term where $x$ does not occur; then the constraint atom  $y=\sharp\{ x\mid \alpha\;\wedge\; t<nx\}$ is equivalent to
 $$
 \bigvee_{l=0}^{n-1}  \exists q\, (t+1= nq+l \;\wedge\; y=\sharp\{ x\mid \alpha\;\wedge\; q-1<x\})~~.
 $$
\end{lemma}

\noindent
\textit{Proof.}  The same as for the previous lemma.
 $\hfill\dashv$